\documentclass[prl,superscriptaddress,twocolumn,showpacs]{revtex4}
\usepackage{epsfig,multirow}

\begin{document}
\title{Magnetic-field asymmetry of electron wave packet transmission
in bent channels capacitively coupled to a metal gate}
\author{R. Kalina} \affiliation{Faculty of Physics and Applied
Computer Science, AGH University of Science and Technology, al.
Mickiewicza 30, 30-059 Krak\'ow, Poland}
\author{B. Szafran}
\affiliation{Faculty of Physics and Applied Computer Science, AGH
University of Science and Technology, al. Mickiewicza 30, 30-059
Krak\'ow, Poland}
\author{S. Bednarek}
\affiliation{Faculty of Physics and Applied Computer Science, AGH
University of Science and Technology, al. Mickiewicza 30, 30-059
Krak\'ow, Poland}
\author{F.M. Peeters}
\affiliation{Departement Fysica, Universiteit Antwerpen,
Groenenborgerlaan 171, B-2020 Antwerpen, Belgium}

\begin{abstract}
We study the electron wave packet moving through a bent channel. We
demonstrate that the packet transmission probability becomes an
uneven function of the magnetic field when the electron packet is
capacitively coupled to a metal plate. The coupling occurs through a
non-linear potential which translates a different kinetics of the
transport for opposite magnetic field orientations into a different
potential felt by the scattered electron.
\end{abstract}
\pacs{73.40.Gk, 73.63.Nm, 72.20.Ht} \maketitle The current ($I$)
that flows through mesoscopic conductors in contact with electron
reservoirs is expressed as the product of the voltage ($V$) applied
to the leads and the conductance ($G$) which in the linear transport
regime depends on the external magnetic field ($B$) but not on the
voltage,
 $I(B,V)=G(B)V$.
According to the Landauer theory the linear conductance ($G$) of a
two-terminal conductor
 $G(B)=\frac{e^2}{h} T_{12}(B)$,
 is proportional to the transfer probability of a Fermi level
 electron from terminal 1 to terminal 2 ($T_{12}$).
For a conductor with disorder (or in general with a potential
landscape) that is independent of $B$, or is an even function of
$B$, the transmission probability is symmetric with respect to the
magnetic field orientation. The relation
\begin{equation}T_{12}(B)=T_{12}(-B),\end{equation} known as Onsager symmetry when
expressed in terms of
 conductance,
can be derived in the scattering matrix theory \cite{ytat}.
 Time-reversal symmetry alone
gives only $T_{12}(B)=T_{21}(-B)$, which is due to the fact that the
electron trajectory from terminal 1 to 2 deflected by the Lorentz
force for magnetic field $+B$ coincides with the trajectory from 2
to 1 and opposite field orientation $-B$. The symmetry (1) results
from the backscattering probability being an even function of the
magnetic field $R_{11}(B)=R_{11}(-B)$, since the backscattered
trajectories are independent of the magnetic field orientation [see
Fig. 1(a)].
Recently it was pointed out \cite{sb} that a current asymmetry
 $I(B,V)\neq I(-B,V)$ can occur in the non-linear transport
regime, i.e., when the conductor is driven out of equilibrium.
Magnetic-field asymmetry of the non-linear conductance was indeed
found in carbon nanotubes \cite{cn}, quantum billiards \cite{qb}
rings \cite{qr} and dots \cite{qd}.

\begin{figure}[ht!]
\begin{tabular}{cc}
\begin{tabular}{c}
\epsfysize=20mm (a) \epsfbox[30 346 518 571]{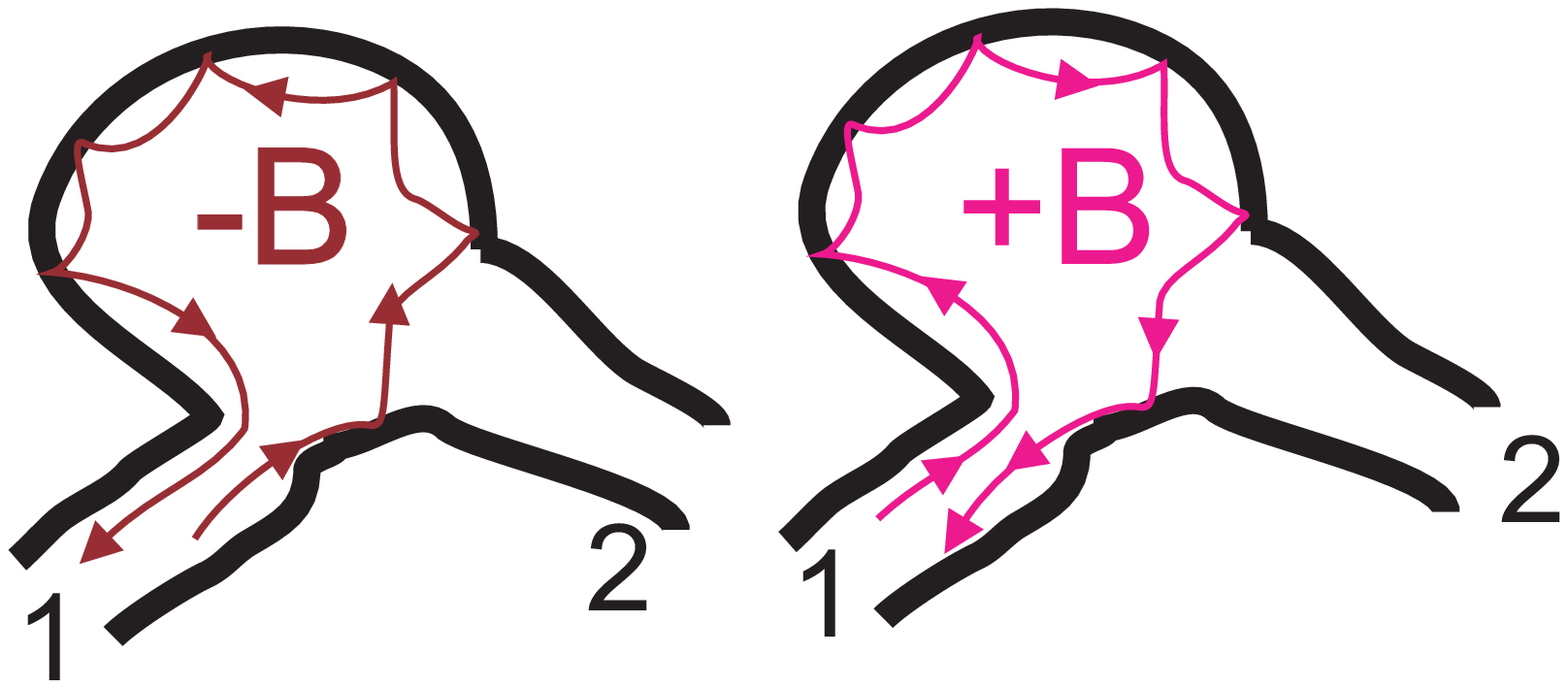} \\
\epsfysize=25mm
                (b) \epsfbox[50 35 544 329]
                {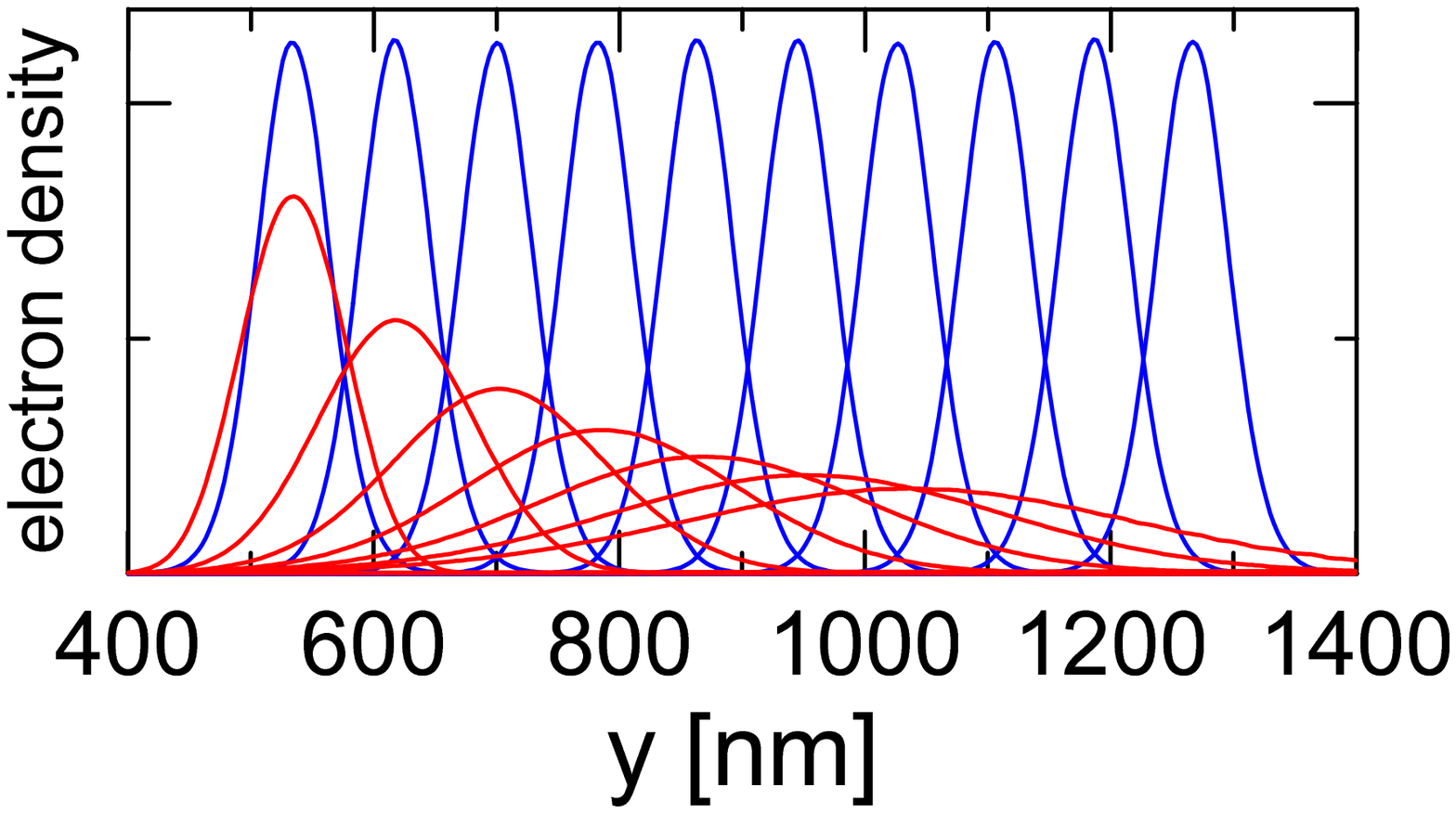}
\end{tabular} &\begin{tabular}{c}
 \epsfysize=40mm(c)\epsfbox[87 110 413 632]
                {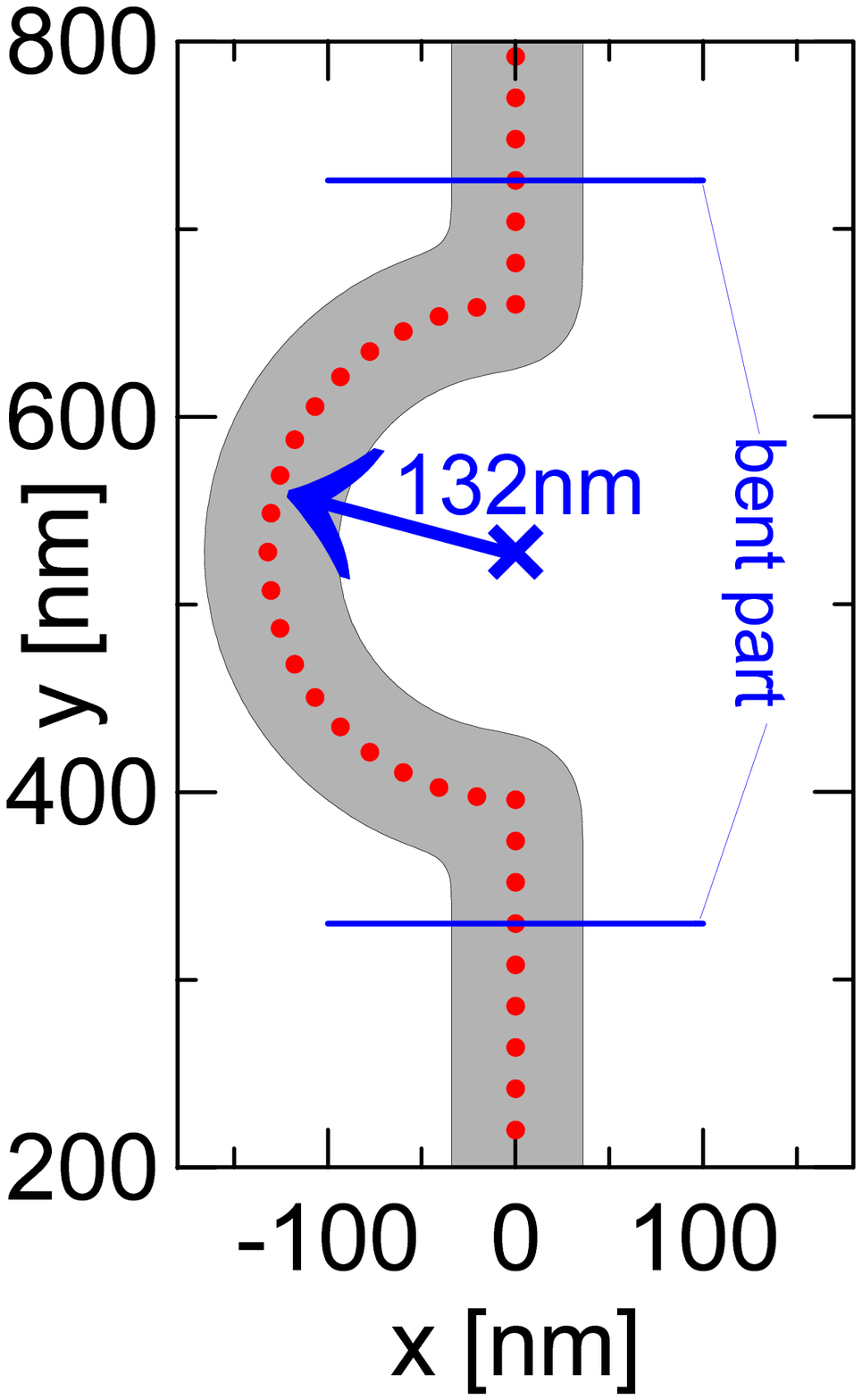}\end{tabular} \\
\end{tabular}
\caption{(color online) (a) Backscattered trajectories for opposite
magnetic field orientations. For $-B$ ($+B$) the trajectories are
deflected to the right (left) by the Lorentz force. (b) Snapshots of
the wave packet density along the axis of the straight wire that is
covered (blue lines) and uncovered (red lines) by a metal plate.  c)
Dots show the position of the centers of the basis function [Eq.
(2)] defining the channel. Shaded area shows the channel range
determined as the region where the sum of the absolute values of the
basis functions is larger than 6.5\% of its maximum value. The
horizontal lines show the range of the bent part $y\in[330 \mathrm{
nm},726\mathrm{ nm}]$.} \label{p1}
\end{figure}

In this Letter we report on simulations of the transport of an
electron packet through a bent semiconductor channel that is
capacitively coupled to a metal plate. In gated semiconductor
nanodevices \cite{gatednano} the confined charge carriers are
capacitively coupled to the electrodes. The coupling is described by
the Schr\"odinger-Poisson scheme \cite{gatednan}. The electron wave
packet induces a redistribution of the charge on the surface of the
metal. The redistributed charge on the metal surface is a source of
potential that focuses the electron wave function and stabilizes the
shape of the packet when it is set in motion \cite{naszprl}. The
electron density interaction with the metal is non-linear and
results in a cancelation of dispersion effects intrinsically present
in the linear Schr\"odinger equation \cite{stare}. For that reason
the electron wave packet interacting with the metal has a soliton
character. We demonstrate below that the transmission probability of
the electron soliton through a bent channel is an uneven function of
the magnetic field.

Our simulation is based on a numerical solution of the
two-dimensional time-dependent Schr\"odinger equation $i\hbar
\frac{\partial \Psi}{\partial t}=H\Psi$. We use an approach
\cite{time} originally introduced to describe the attenuation of the
Aharonov-Bohm oscillations by the preferential injection of the
incoming wave packet to one of the arms of the quantum ring by the
Lorentz force. The kinetic energy Hamiltonian is given by
$T=\left(-i\hbar\nabla+e{\mathbf A}\right)^2/2m^*$. The Landau gauge
${\mathbf A}=(-By,0,0)$, GaAs electron effective mass $m^*=0.067m_0$
and dielectric constant $\epsilon = 12.4$ are used. The time
evolution is calculated in the basis of Gaussian functions
$\Psi({\bf r},t)=\sum_n c_n(t) f_n({\bf r})$, where
\begin{equation} f_n({\bf r})= \exp\{-\alpha({\mathbf
r}-\mathbf{R}_n)^2 +  i\beta (x-X_n)(y+Y_n)\}, \label{fn}
\end{equation}
with ${\bf r}=(x,y)$, $\beta=eB/2\hbar$, $\alpha=1/(22$nm$)^2$ and
$\mathbf{R}_n=(X_n,Y_n)$  is the center of the $n$-th Gaussian. The
imaginary term in (2) is related to the magnetic translation and
ensures gauge invariance. In the applied approach the centers
$\mathbf{R}_n$ define the structure itself. The choice modeling the
bent channel is shown in Fig. 1(c).

We assume that at a distance $d=20$ nm above the channel the
structure is covered by an infinite metal plane. For infinite plane
the solution of the Poisson equation \cite{naszprl} for the response
of the metal can be obtained with the image-charge technique
\cite{stare}. Interaction energy of the electron density with metal
introduces is therefore given by
\begin{equation}
V({\bf r},t)=\int  \frac{-|e\Psi({\bf r}',t)|^2d{\bf
r}'}{{4\pi\epsilon\epsilon_0}\left[(x-x')^2+(y-y')^2+(2d)^2\right]^{1/2}}.
\end{equation}
All the time dependence of the wave function (2) is introduced in
the expansion coefficients $c_n(t)$. Equations for $c_n(t)$ are
obtained from the Crank-Nicolson differential scheme $i\hbar
\left[\Psi(t+\Delta t)-\Psi(t)\right]/\Delta t =\left[H(t+\Delta
t)\Psi(t+\Delta t)+H(t)\Psi(t)\right]/2$. The Hamiltonian depends on
time through the induced potential $H(t)=T+V(t)$. For the basis (2)
the Crank-Nicolson scheme is given by a
system of equations $\textbf{Sc}(t + \Delta t) = \textbf{Sc}(t) -
\frac{i\Delta t}{{2\hbar }}\left[ {\textbf{H}(t + \Delta
t)\textbf{c}(t + \Delta t) + \textbf{H}(t)\textbf{c}(t)} \right]$,
where the overlap and Hamiltonian matrix elements are given by
$\textbf{S}_{mn}=\left<f_m|f_n\right>$ and
$\textbf{H}_{mn}(t)=\left<f_m|H(t)|f_n\right>$, respectively. As the
initial condition we take a self-focused wave packet \cite{stare}
localized 500 nm before the bent segment moving \cite{time} along
the incoming wire with average kinetic energy $\hbar^2 k^2/2m^*=1.4$
meV ($k=0.05/$nm).

%

\begin{figure}[ht!]
\hbox{\epsfysize=110mm
               \epsfbox[63 100 518 800] {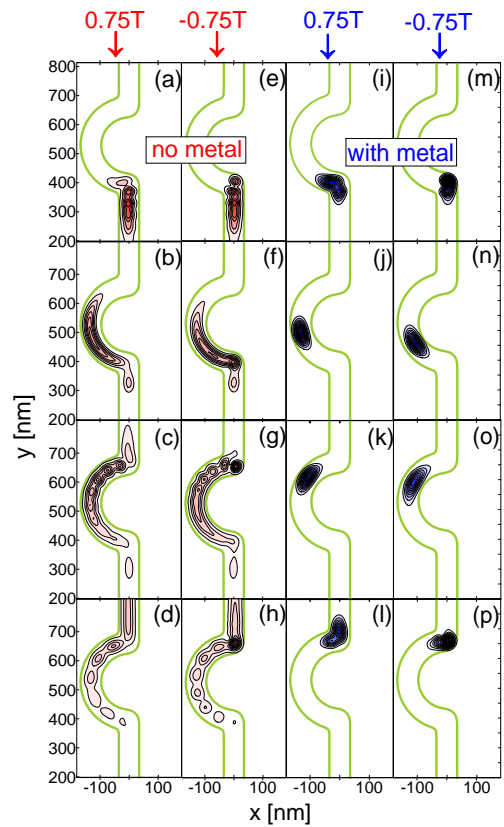}\hfill}
\caption{(color online) Snapshots of the motion of the electron
packet injected into the bent wire from the lower lead for opposite
magnetic field orientations. Later moments in time correspond to
lower plots. Solution of the linear Schr\"odinger equation (a-h) is
presented for $t=2.1, 4.8, 6.7$ and 9.5 ps and the solution
including the non-linear term of the wave packet-metal interaction
(i-p) is given for $t=3.4, 5.8, 7.7$ and 9.7 ps. Green lines delimit
the channel as defined in Fig. \ref{p1}(c).
 \label{pot}}
\end{figure}

\begin{figure}[ht!]
\hbox{\epsfysize=80mm
               \epsfbox[98 290 525 730] {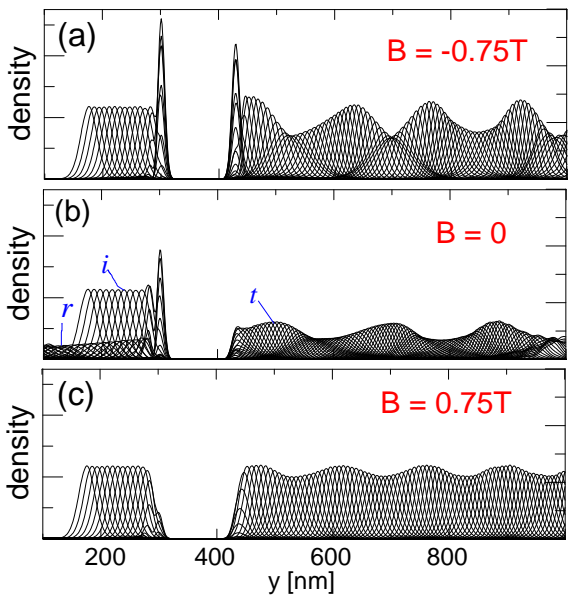}\hfill}
\caption{(color online) The snapshots of the packet density along
the axis of the straight channels ($x=0$) for the electron
interacting with metal for $B=-0.75$ T, 0 and $0.75$ T in (a), (b)
and (c), respectively. In (b) the incoming, reflected and
transmitted wave packets are marked by {\it i, r} and {\it t},
respectively.}
\end{figure}

\begin{figure}[ht!]
\hbox{\epsfysize=35mm
               (a)\epsfbox[77 586 511 824] {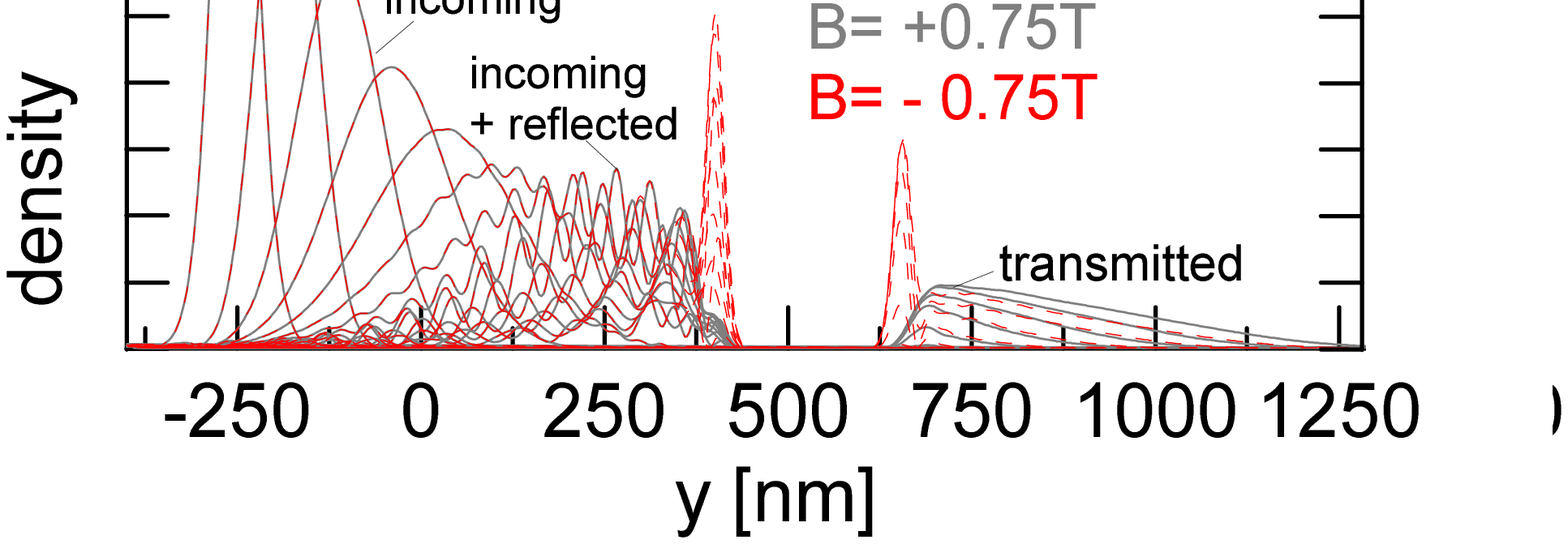}\hfill}
               \hbox{\epsfysize=35mm
               (b)\epsfbox[114 586 547 824] {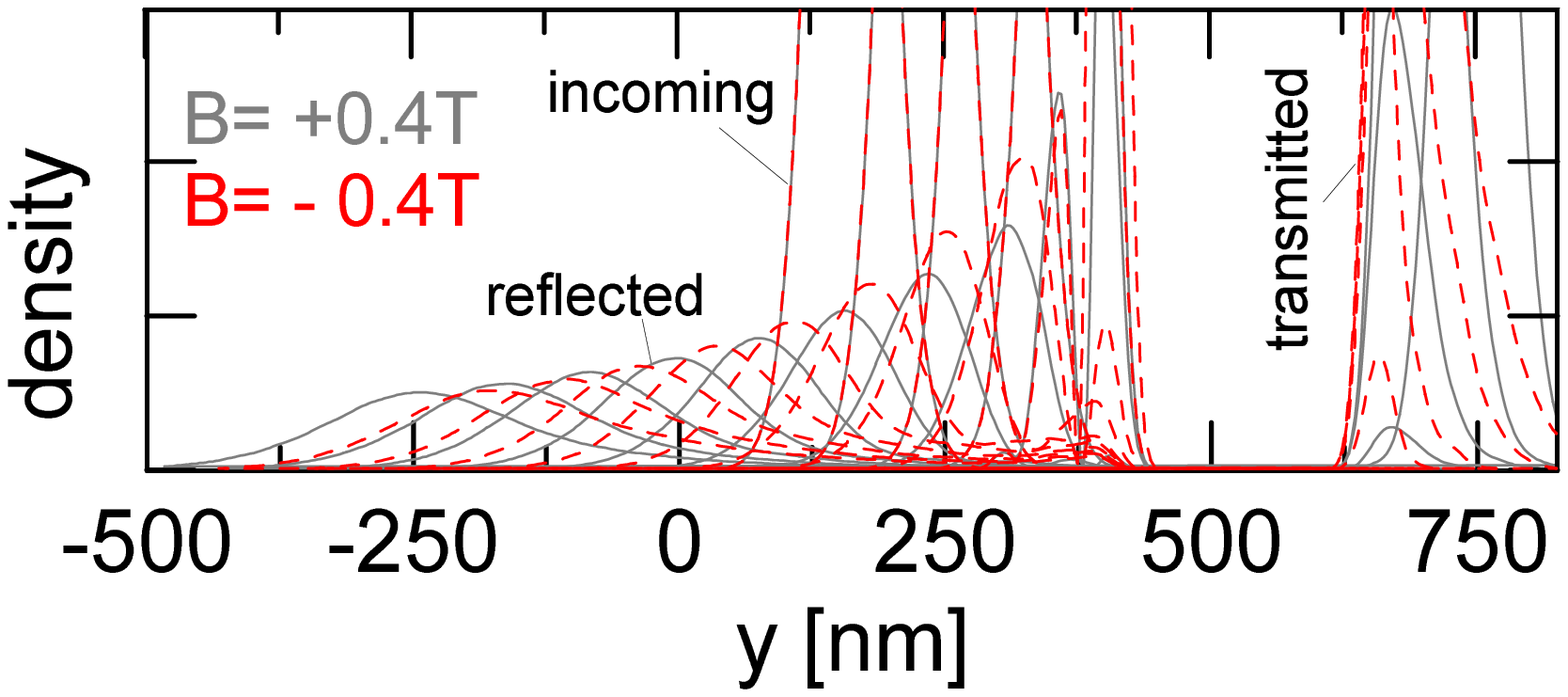}\hfill}
\caption{(color online) Same as Fig. 3 for the normal wave packet
(a) and a wave packet interacting with the metal plate (b).}
\end{figure}

\begin{figure*}[ht!]
\hbox{
               (a)\epsfysize=55mm\epsfbox[23 46 553 573] {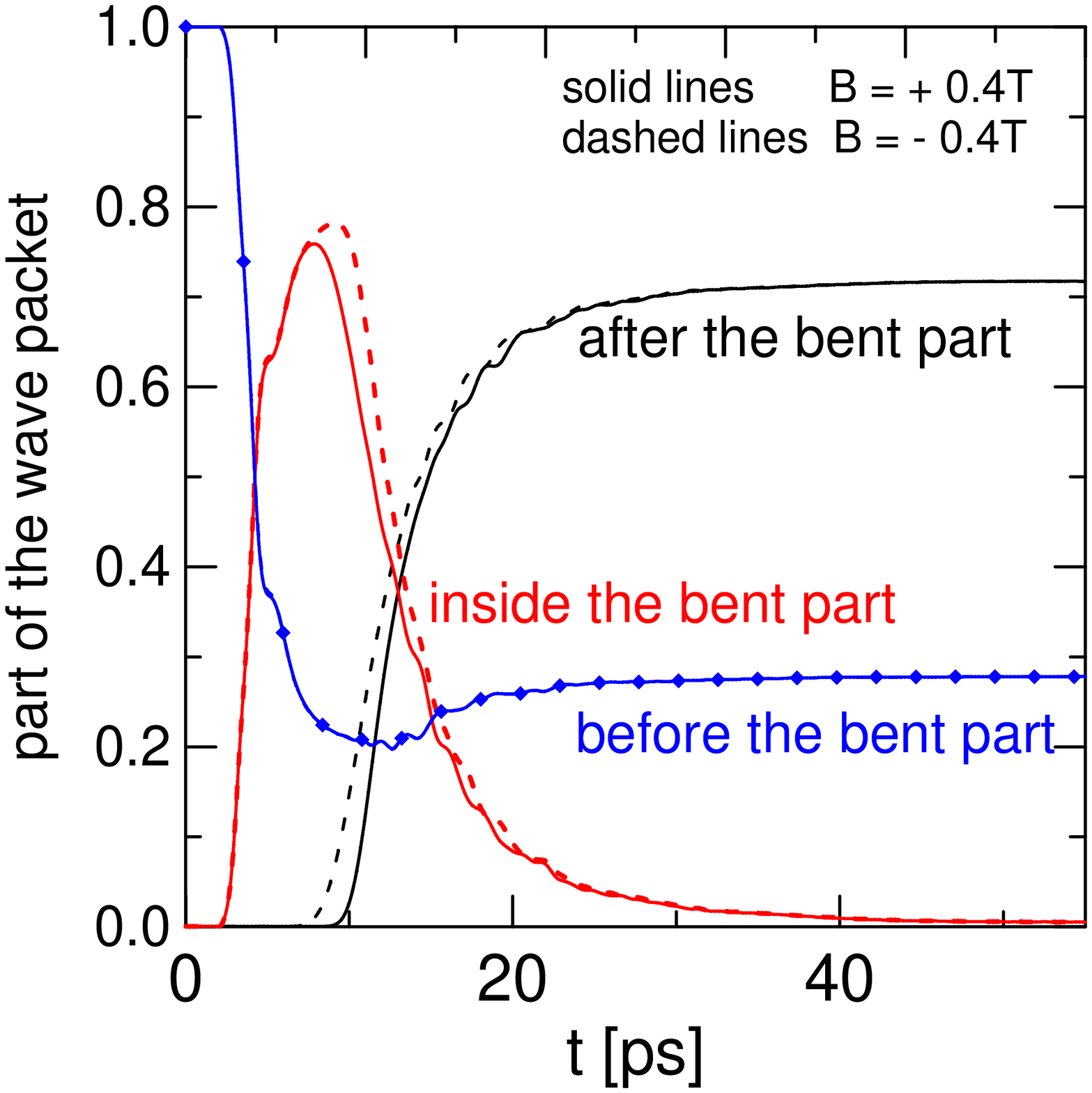}

               (b)\epsfysize=55mm\epsfbox[23 46 553 573]
               {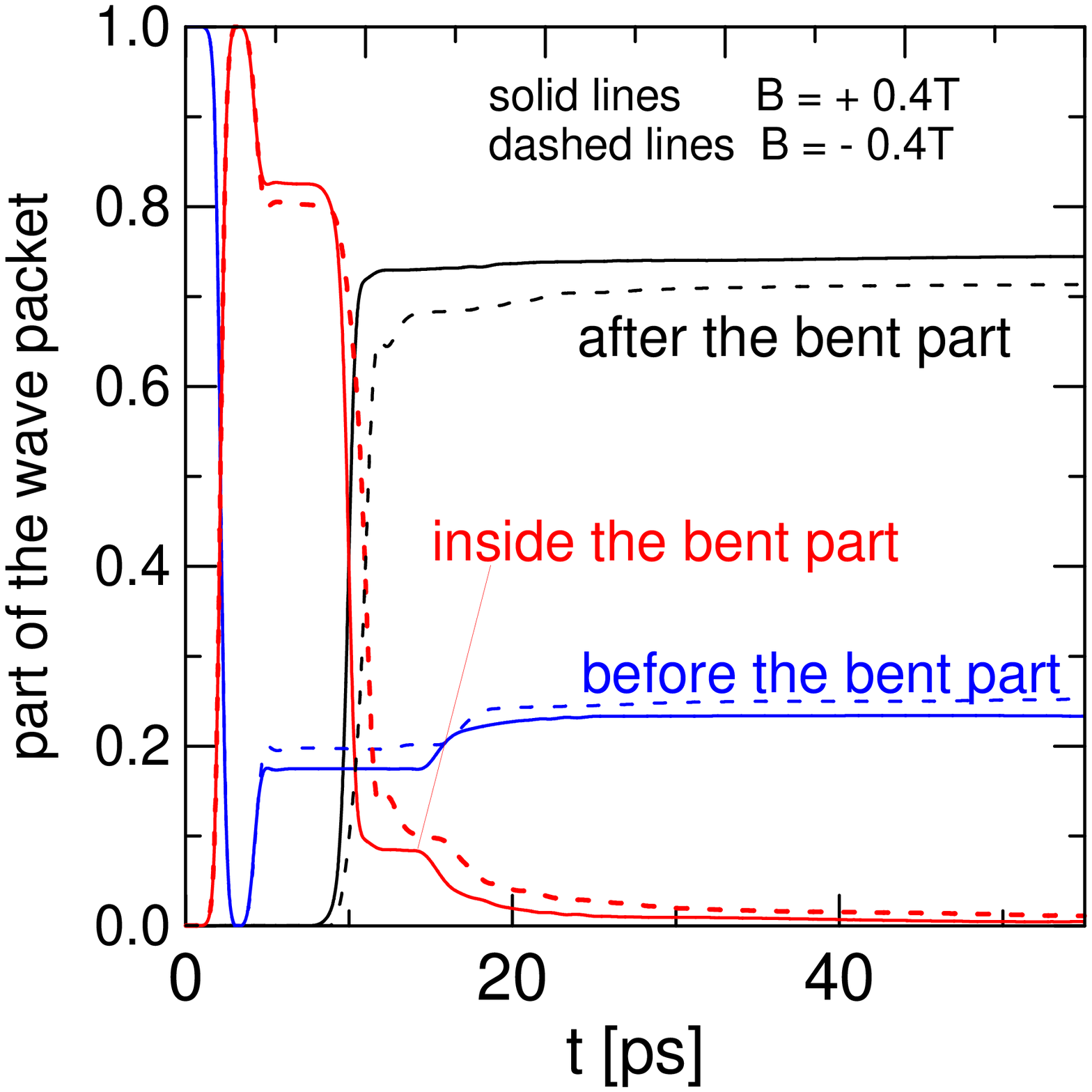}
               (c)\epsfysize=55mm\epsfbox[160 490 411  753] {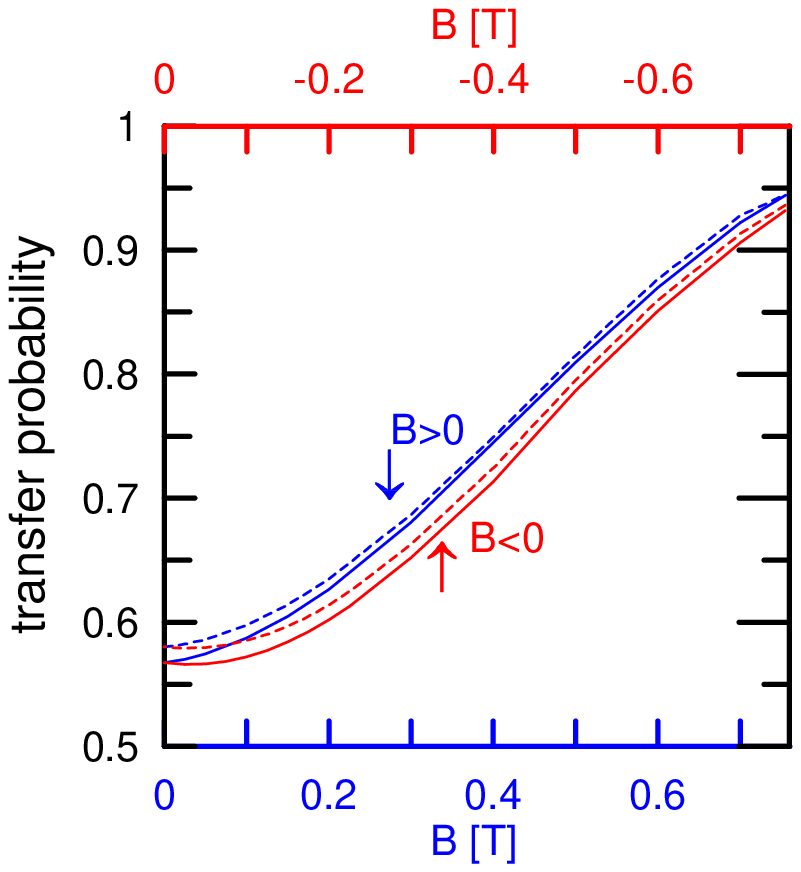}\hfill}
\caption{(color online) Parts of the wave packet before, after and
inside the bent part of the wire for the electron density uncoupled
(a) and coupled (b) to the metal plate. Solid lines show the results
for $B>0$ and dashed lines for $B<0$. In (a) the part 'before' the
packet for $B<0$ is given by the symbols. (c) Lower (solid) and
upper (dashed) curves are bounds for the wave packet transfer
probability for $B>0$ (blue lines) and $B<0$ (red lines).
 \label{czas}}
\end{figure*}

A comparison of the motion of the soliton and normal (i.e.
non-interacting with metal) wave packet moving in a straight channel
is shown in Fig. 1(b). Scattering on the bent segment is illustrated
in Fig. 2 which shows the density of the moving packet at specific
times (increasing from top to bottom figures) for the situation
without (two left columns) and with a metal gate present (two right
columns) for $B=\pm 0.75$T. For positive value of the field the wave
packet is deflected to the left by the Lorentz force as it moves up
[see also Fig. 1(b)] and thus it is injected into the bent of the
wire [Figs. 2(a,e)]. For negative $B$ value the injection occurs
only after the velocity of the packet is inverted [Fig. 2(e,m)] when
the packet is reflected at the entrance. This introduces a delay in
the transport of the packet for negative $B$ value.


More details of the motion are seen on the plots of the density
taken along the straight part of the wires ($x=0$). For $B=-0.75$ T
we notice [Fig. 3(a)] a compression of the wave packet which occurs
when the packet is reflected at the entrance and at the exit of the
bent part. No compression is seen for $B=0.75$ T, for which the
Lorentz force guides the packet into the bent and out of it without
reflection at the junctions. For both $B=\pm 0.75$ T the wave packet
transfer probability is nearly 95\% and no reflected part of the
packet can be noticed at this scale of the plot. The reflected part
is visible for zero magnetic field [Fig. 3(b)]. In all cases the
transmitted packet stays more or less compact, but
 with a shape that changes in time. The breathing of the transmitted wave packet
 is due to deviations of its form from the stable one
that occurs during the transfer through the bent segment. For
$B=+0.75$ T the breathing is nearly absent, and it is strongest for
$B=-0.75$ T for which wave packet compression at the junctions
occurs. The results for the normal packet are given in Fig. 4(a)
where the densities are calculated at the same moments in time for
both field orientations. Both densities in the straight wire are
exactly the same in the region before the bent part for the incoming
wave packet as well as for the interference which occurs when slower
(still incoming) and faster (already reflected) parts of the wave
packet meet below the bent. Note that such an interference was not
observed for the soliton wave packet. Differences in the shape of
the packet for $B=\pm 0.75$ T are clearly visible at the entrance to
the bent  (compression for negative $B$) and in the transmitted wave
packet. The reflected part of the soliton scattering  for $B=\pm
0.4$ T is observed in Fig. 4(b). Only the plots for the incoming
wave packet coincide for both $B$ orientations. The snapshots of the
reflected parts of the wave packet are no longer the same when the
magnetic field is inverted. We observe here a destruction of the
soliton which is split into two parts. Note, that the reflected part
of the scattered soliton [see Fig. 4(b)] spreads like a normal wave
packet. The image charge of this part is not large enough to focus
it.

The time-dependence of the fraction of the wave packet inside the
bent as well as in the straight wires before and after the bent [see
Fig. 1(c)] are shown in Fig. 5(a) for the normal packet and in Fig.
5(b) for the soliton. In the absence of the metal the fraction of
the wave packet before the bent is exactly the same for both $\pm
0.4$ T. Since the bent segment does not permanently capture any part
of the normal packet the fraction of the packet which is eventually
transmitted to the outgoing wire is the same for both $B$
orientations. For the wave packet interacting with the metal [see
Fig. 5(c)] the reflected part of the wave packet before the bent
segment is different for $\pm 0.4$ T and a different transmission
probability is obtained. Soliton destruction is inelastic and
associated with decreased absolute value of both electron-metal
interaction energy and the kinetic energy of the packet.

The fraction of the wave packet which is transmitted into the
outgoing wire at the end of the simulation is a {\it lower bound} to
the transmission probability. A small fraction of the wave packet
still bounces back and forth within the bent leaving it in small
portions to the incoming and to the outgoing wires. For that reason
the sum of the transmitted wave packet and the part which is still
present in the bent at the end of the simulation is an {\it upper
bound} to the transmission probability. The bounds on $T(B)$ for the
wave packet interacting with the metal plate are plotted in Fig.
5(c). We conclude that the transmission probability is: i) an uneven
function of the magnetic field, and ii) for $B>0.1$ T the asymmetry
in the transmission probability for opposite $B$ orientations
exceeds the evaluation accuracy.

In the absence of the metal plate the fraction of the normal packet
in the incoming lead $R(B,t)$ is an even function of the magnetic
field at any moment in time. In general also for the normal packet
the transferred fraction is different for opposite field
orientations, i.e. $T(B,t)\neq T(-B,t)$, and a symmetric transfer
probability is obtained only in the large $t$ limit. The kinetics of
the transfer of both normal and solitary packets through an
asymmetric channel is different for opposite field orientations. The
different kinetics for the wave packet interacting with the metal is
translated into a different potential felt by the scattered
electron. For that reason the argument of the equivalent
backscattered trajectories for opposite $B$ orientations [cf. Fig.
1(a)] does not hold in the present case. The effective scattering
potential is different when the magnetic field is inverted and
consequently $R(B,t)\neq R(-B,t)$, also in the large $t$ limit. The
reason of the magnetic-asymmetry is the asymmetry of the inelastic
soliton destruction for $\pm B$.

To further convince ourselves, we studied the case of the electron
dynamics governed by the Gross-Pitaevski equation \cite{GP} in which
the potential (3) is replaced by $V({\bf r},t)=-\gamma |\Psi({\bf
r},t)|^2$ with $\gamma>0$ (obtained for $1/r_{12}$ replaced by Dirac
$\delta(r_{12})$ in the integrand of (3)).
 For this potential the Hamiltonian matrix
elements can be evaluated analytically. We found that also in this
case the transmission probability is asymmetric in $B$.

The presented theory was applied to an empty ballistic channel that
is capacitively coupled to a metal plate with a carefully prepared
initial wave packet. The channel is not connected to electron
reservoirs and thus the modeling does not correspond to an electron
gas in equilibrium. Although the theory does not allow us to
calculate the current, it is nevertheless related to the previous
work on the magnetic-field asymmetry of non-linear conductance. In
the original idea \cite{sb} it was the dependence of the potential
landscape on $B$ that resulted in the asymmetry. In our model the
role of the landscape is played by the response potential of the
metal which depends on $B$ orientation through the time dependence
of the image charge. In previous work \cite{ci} it was pointed out
that the uneven modification of the potential landscape and the
resulting magneto-asymmetry of non-linear transport is due to the
Coulomb interaction. In our case the moving electron wave packet
interacts with the electrons in metal. It was recently explained
that the Onsager symmetry is broken for conductors interacting with
the environment \cite{sk}. A second conductor \cite{sk} was used as
the environment which was driven out of equilibrium by the applied
bias. In the present work the metal is perturbed by the moving wave
packet.

In conclusion, we demonstrated that the transfer probability of an
electron wave packet through a semi-circular channel in interaction
with a metal plate, that leads to a non-linear potential, is not an
even function of the magnetic field. 

{\it Acknowledgements.} This work was supported by the EU NoE
SANDiE, the Belgian Science Policy (IAP) and the Flemish Science
Foundation.

\end{document}